\documentclass[runningheads]{llncs}
\usepackage[T1]{fontenc}
\usepackage[utf8]{inputenc} % allow utf-8 input
\usepackage{hyperref}       % hyperlinks
\usepackage{url}            % simple URL typesetting
\usepackage{booktabs}       % professional-quality tables
\usepackage{amsfonts}       % blackboard math symbols
\usepackage{nicefrac}       % compact symbols for 1/2, etc.
\usepackage{microtype}      % microtypography
\usepackage{xcolor}         % colors
\usepackage{graphicx}
\usepackage{import}
\usepackage{enumitem,amssymb}
\usepackage{subcaption}
\usepackage{fancyhdr}

\usepackage{macros}
\begin{document}
\title{Revisiting Variable Ordering for Real Quantifier Elimination using Machine Learning}
\titlerunning{Revisiting Variable Ordering for RQE using Machine Learning}
\pagestyle{fancy}
\fancyhead{} % clear all header fields
\renewcommand{\headrulewidth}{0pt}
\renewcommand{\footrulewidth}{.5pt}
\fancyhead[LE]{\thepage~\quad~Hester et al.}
\fancyhead[RO]{Revisiting Variable Ordering for RQE using Machine Learning~\quad~\thepage}
\fancyfoot[C]{\footnotesize Distribution Statement A – Approved for Public Release, Distribution Unlimited}
\author{
    John Hester\inst{1}\thanks{Authors ordered alphabetically.\\This research was, in part, developed with funding from the Defense Advanced Research Projects Agency (DARPA). The views and conclusions contained in this document are those of the authors and should not be interpreted as representing the official policies, either expressed or implied, of the U.S. Government or DARPA.} \and 
    Briland Hitaj\inst{2} \and
    Grant Passmore\inst{1} \and
    Sam Owre\inst{2} \and
    \\Natarajan Shankar\inst{2} \and
    Eric Yeh\inst{2}
}
\authorrunning{Hester et al.}

%First names are abbreviated in the running head.
%If there are more than two authors, 'et al.' is used.

\institute{
    Imandra Inc., Austin, TX 78704, USA\\
    \email{\{john,grant\}@imandra.ai}\\
    % \url{https://www.imandra.ai/} 
    \and
    SRI International, Menlo Park, CA 94025, USA
    \email{\{briland.hitaj,natarajan.shankar,sam.owre,eric.yeh\}@sri.com}\\
    % \url{https://www.sri.com/}
}

% \author{Anonymous submission}
% \authorrunning{Anonymous submission}

%
\maketitle              % typeset the header of the contribution
% <sections>
% !TEX root = ../main.tex
\begin{abstract}
% The abstract should briefly summarize the contents of the paper in
Cylindrical Algebraic Decomposition (CAD) is a key proof technique 
for formal verification of cyber-physical systems. 
CAD is computationally expensive, with worst-case doubly-exponential complexity. 
Selecting an optimal variable ordering is paramount to efficient use of CAD. 
Prior work has demonstrated that machine learning can be useful in determining efficient variable orderings. 
Much of this work has been driven by CAD problems extracted from applications of the MetiTarski theorem prover.
In this paper, we revisit this prior work and consider issues of bias in existing training and test data. 
We observe that the classical MetiTarski benchmarks are heavily biased towards particular variable orderings. 
To address this, we apply symmetries to create a new dataset containing more than 41K MetiTarski challenges designed to remove bias. 
Furthermore, we evaluate issues of information leakage, and test the generalizability of our models on the new dataset. 

\end{abstract}
% !TEX root = ../main.tex
\section{Introduction}
\label{sec:introduction}

Cylindrical Algebraic Decomposition (CAD) is a key proof technique for formal verification of cyber-physical systems such as aircraft collision avoidance systems, autonomous vehicles and medical robotics. 
While CAD is a complete decision procedure, it is computationally expensive with worst-case exponential complexity. 
However, the expense of running CAD on a given problem can be greatly affected by the way in which the problem is posed.
In particular, selecting an optimal variable ordering is paramount to efficient use of CAD. 
Prior work has demonstrated that machine learning (ML) may be fruitfully applied to determining efficient variable orderings~\cite{england2019comparing,huang2019using}. 
Much of this work has been driven by CAD problems extracted from applications of the MetiTarski theorem prover~\cite{MetiTarski2010JAR,passmore2012real}.

Generalizability is a major concern for ML-trained systems~\cite{geirhos2018generalisation,kawaguchi2017generalization}.  For example, in visual object classification systems, if objects to classify in training sets always occur in the upper left of an image, the classifier may fail if it encounters objects in the bottom right.

The identification of symmetries is a major factor in creating robust ML systems.  Symmetries are transformations of data that preserve relevant semantics.  Image based object classification provides many natural examples, where transformations such as mirror flipping or large random crops do not change the identity of the object in the image.  These symmetries form the basis of data augmentation methods, where such transformations applied over existing training data can create additional training data.  This augmentation can substantially improve the robustness of trained systems~\cite{geirhos2018imagenettrained}.

We argue that this problem also exists in applications to mathematical tasks such as CAD variable ordering.  In this case, variables may be permuted over both statements and target orderings while preserving semantics.  Performing an augmentation that debiases using formula symmetries is important for maintaining performance in a statistically trained ML system.  

% \subsection{Contributions}
\subsubsection{Contributions:}
%
% Our contributions may be summarized as follows:
%
\begin{itemize}
    \item We observe bias in the classical MetiTarski benchmarks often used in training and evaluating ML-based CAD variable ordering selection systems.
    \item We show how data augmentation along inherent symmetries can remove bias and improve generalizability, resulting in a new ``debiased'' benchmark set.
    \item We argue that this phenomenon is more general than CAD, and that debiasing using formula symmetries should be a standard tool for applications of ML in computer algebra, program verification, and other fields manipulating mathematical formulas.
\end{itemize}
% !TEX root = ../main.tex
\section{The CAD Variable Ordering Problem}
\label{sec:background}

\subsection{CAD-based Real Quantifier Elimination}

Consider a system of nonlinear equations and inequalities over $\mathbb{R}$:
\[
  \varphi \  =  \ \left( \bigwedge_{i=1}^k p_i(x_1, \mathellipsis, x_n) \odot_i 0 \right), \ p_i \in \mathbb{Q}[x_1, \mathellipsis, x_n], \ \odot_i \in \{ <, \leq, =, \geq, > \}.
\]
CAD may be used to decide the satisfiability of this system, i.e., 
the truth of 
\[
\exists x_1, \mathellipsis, \exists x_n \in \mathbb{R}. \ \varphi(x_1, \mathellipsis, x_n) 
\]
by 
(a) partitioning $\mathbb{R}^n$ into finitely many connected regions called \emph{cells}
$C = \{c_1, \mathellipsis, c_m\} \subseteq \mathbb{R}^n$ s.t. each $p_i$ is
\emph{sign-invariant} over each $c_j$, and 
(b) evaluating $\varphi$ at a sample point $s_j \in c_j$ drawn from each cell.
Given sign-invariance and the fact that $C$ is a covering ($\bigcup C = \mathbb{R}$), 
it follows that $\varphi$ is satisfiable over $\mathbb{R}^n$ iff it evaluates to true on at least
one such sample point. This approach can be readily extended to formulas involving quantifier
alternation~\cite{collins1975cad,Passmore2011PhDthesis}.

\subsection{Efficient Variable Ordering}

CAD construction proceeds in stages, eliminating each variable $x_i$ in turn. The
\emph{variable ordering} is the order in which variables are eliminated. In the present work, we are concerned with determining an efficient variable ordering for a given problem.
For example, consider a CAD induced by polynomials $P = 
\{68x_1^2 - 12x_3x_2 + 46x_3 - 126, -54x_2x_1 + 11x_1 + 92x_2 - x_3x_1^2 - 42x_3x_2x_1 + 45x_4^2 - 35\}$ over $\mathbb{Q}[x_1,x_2,x_3,x_4]$. Depending on the variable ordering, the efficiency of CAD computation can vary substantially~\cite{chen2020varord}. For example, with ordering $x_4 \succ x_3 \succ x_2 \succ x_1$ a CAD is constructed in 5s with 3,373 cells, while with $x_3 \succ x_1 \succ x_4 \succ x_2$ a CAD is constructed in 93s with 43,235 cells.

% \import{\sectiondir}{03-related-work.tex}
% !TEX root = ../main.tex
\section{Experimental Setup}
\label{sec:experimental-setup}

% \briland{Details about the datasets used/generated, machine learning architectures, hyperparameter setup, hardware}
In this section, we provide details about the datasets that we have used in our experiments, and we introduce a new \emph{augmented} dataset designed to remove bias. In addition to the datasets, we provide details on the features considered based on England et al.~\cite{england2019comparing} and Huang et al.~\cite{huang2019using} respective works, followed by a discussion of our labeling strategy. We conclude the section with a discussion of the machine learning models considered together with their respective hyperparameter setup.

\subsection{MetiTarski Datasets}
\label{ssec:metitarski-datasets}

\subsubsection{Dataset 1 (Original):}
The first dataset is predominantly gathered from MetiTarski, by logging MetiTarski's RCF subproblem queries during its proof search~\cite{passmore2012real,england2019comparing}.
The dataset contains 6,895 polynomial systems, together with data on the performance of a CAD-based decision procedure on different orderings of their variables. Every problem in this data set has 3 variables ($x_1, x_2, x_3$) and thus 6 possible variable orderings.
%, Figure~\ref{fig:polynomial_output_100}.

\subsubsection{Dataset 2 (Augmented):} %41370
\begin{figure}[]
     \centering
     \begin{subfigure}[b]{0.45\textwidth}
         \centering
         \includegraphics[width=\textwidth]{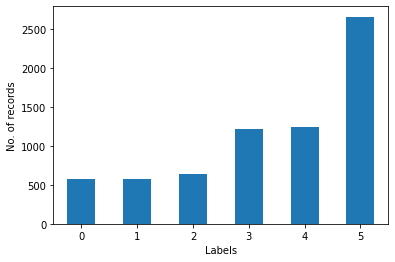}
         \caption{Original Dataset}
         \label{fig:metitarski-original}
     \end{subfigure}
     \hfill
     \begin{subfigure}[b]{0.45\textwidth}
         \centering
         \includegraphics[width=\textwidth]{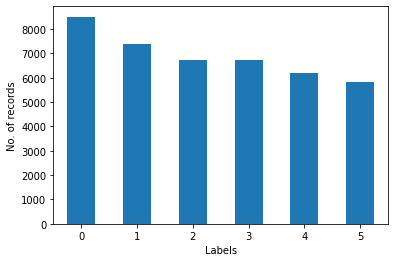}
         \caption{Augmented Dataset}
         \label{fig:metitarski-balanced}
     \end{subfigure}
        \caption{Data distribution per label on both the original MetiTarski dataset and the second, balanced one.}
        \label{fig:metitarski-data-distribution}
\end{figure}

We noted that the original MetiTarski dataset was highly imbalanced. While class $0$, corresponding to the ($x_1, x_2, x_3$) variable order contained 580-records, class $5$, corresponding to the ($x_3, x_2, x_1$) variable order contained 2,657-records, nearly 4-times more, Figure~\ref{fig:metitarski-original}. 
We note that this may lead ML models to be \emph{biased} towards certain label/s, thus preventing the models to learn relevant information and hinder their generalizability to new, previously unseen data samples.%~\cite{}. 

We recognize that variables in the formula and ordering can be swapped without changing the time and cost needed to perform the computation.  For instance, swapping variables $x_1$ with $x_2$ in the formulas and in the ordering leads to a CAD with the same time and cell cost.  While this may seem apparent to a human or a machine reasoner that already has the ability to recognize this symmetry, for current machine learning systems it needs to be made explicit in the training data. This procedure resulted in a new augmented composed of 41,370 polynomial systems, Figure~\ref{fig:metitarski-balanced}.

\subsection{Feature Engineering and Labeling}
\label{ssec:feature-engineering}

\subsubsection{Feature Engineering:}
We process each set of polynomials extracting features enlisted in~\cite{england2019comparing,huang2019using}, including the number of polynomials, maximum total degree of polynomials, maximum degree of each $x_i$, and proportion of each $x_i$ appearing in polynomials and monomials. 

\subsubsection{Data Labeling:}
In addition to the feature set, we assign to each polynomial problem a label ranging from $0 ... 5$, where each label translates to one of the $6$-possible variable orderings. At present, the label for each polynomial problem corresponds to the variable ordering that takes the least amount of time. 

\subsection{Models}
\label{ssec:models}

To evaluate our approach, we used $5$-ML classifiers: 1) Support Vector Machines (SVM), 2) k-Nearest Neighbours (k-NN), 3) Decision Trees (DT), 4) Random Forests (RF), and 5) Multi-Layer Perceptrons (MLP). In their work, England et al.~\cite{england2019comparing} rely on SVMs, k-NNs, DTs, and MLPs. By following a similar approach, we ensure that our strategy is comparable to the state-of-the-art and thus can be used as a foundation for future adoption of more complex ML strategies, such as Transformers~\cite{vaswani2017attention} or Graph Neural Networks (GNNs)~\cite{Scarselli2009GNNs}.
We used the \texttt{scikit-learn}\footnote{\url{https://scikit-learn.org/stable/}}-based implementations of the ML algorithms. Similar to the works of England et al.~\cite{england2019comparing} and Huang et al.~\cite{huang2019using}, we employed a grid-search strategy with 5-fold cross-validation to identify the right parameter setup for each of the models.

% !TEX root = ../main.tex
\section{Evaluation}
\label{sec:evaluation}

% \briland{TO DO: Add plots and numbers from the Colab notebook}
In this section, we proceed with the evaluation of the performance of the selected machine learning models (cf. Section~\ref{ssec:models}) on identifying the preferred (best) variable order for a given input problem. We transform the problem of determining the best variable order into a multi-class classification problem. In these kind of problems the training set is composed of $(x, y)$ tuples of data, where $x$ is an input sample and $y$ is the corresponding label for that sample. The goal of the learning process translates into the task of finding a function $f$, such that $f(x) = y$. 

In our setting, the input data corresponds to the series of 11-features (cf.  Section~\ref{ssec:feature-engineering}) whereas $y$ is one of the 6-labels from $[0,\mathellipsis,5]$ belonging to a preferred variable order from $[(x_1, x_2, x_3),\mathellipsis,(x_3, x_2, x_1)]$. The features were scaled by subtracting the mean and then scaling to unit variance\footnote{\url{https://scikit-learn.org/stable/modules/generated/sklearn.preprocessing.StandardScaler.html}}.

\subsubsection{Training:}
For each of the datasets, we used $80\%$ of the data for training and kept the remaining $20\%$ as part of the testing set. For Dataset 1 (original), this corresponded to $5,516$ samples for training and $1,379$ for testing, whereas for Dataset 2 (augmented), $33,095$ samples were used during training and the remaining $8,274$ samples for testing. Tables~\ref{tab:performance-dataset-1} and~\ref{tab:performance-dataset-2} provide accuracy of each model obtained on the respective training set.

\begin{table}[]
% \tiny
\centering
\caption{Performance of different models trained on Dataset 1 when evaluated on testing set 1 and entire Dataset 2.}
\label{tab:performance-dataset-1}
% \resizebox{\textwidth}{!}{%
\begin{tabular}{c|c|c|c}
% \hline
\multicolumn{1}{c|}{\textbf{\begin{tabular}[c]{@{}c@{}}Model Trained on\\ Dataset 1\end{tabular}}} & \multicolumn{1}{c|}{\textbf{\begin{tabular}[c]{@{}c@{}}Performance on\\ Training Set 1\end{tabular}}} & \multicolumn{1}{c|}{\textbf{\begin{tabular}[c]{@{}c@{}}Performance on\\ Testing Set 1\end{tabular}}} & \multicolumn{1}{c}{\textbf{\begin{tabular}[c]{@{}c@{}}Performance on\\ Dataset 2 (all)\end{tabular}}} \\ \hline
SVM & 69.38\% & 58.88\% & 28.9\% \\ \hline
k-NN & 75.27\% & 57.36\% & 32.21\% \\ \hline
DT & 75\%\% & 55.69\% & 31.44\% \\ \hline
RF & 76.39\% & 58.23\% & 34.15\% \\ \hline
MLP & 58.81\% & 53\% & 33.64\% \\ \hline
\end{tabular}%
% }
\end{table}

\subsubsection{Testing:}
We test each of the trained models on 20\% of the respective datasets, i.e., on $1,379$ samples for models trained on Dataset 1 and $8,274$ samples for models trained on Dataset 2. 

\begin{table}[]
% \tiny
\centering
\caption{Performance of different models trained on Dataset 2 when evaluated on testing set 2 and entire Dataset 1.}
\label{tab:performance-dataset-2}
% \resizebox{\textwidth}{!}{%
\begin{tabular}{c|c|c|c}
% \hline
\multicolumn{1}{c|}{\textbf{\begin{tabular}[c]{@{}c@{}}Model Trained on\\ Dataset 2\end{tabular}}} & \multicolumn{1}{c|}{\textbf{\begin{tabular}[c]{@{}c@{}}Performance on\\ Training Set 2\end{tabular}}} & \multicolumn{1}{c|}{\textbf{\begin{tabular}[c]{@{}c@{}}Performance on\\ Testing Set 2\end{tabular}}} & \multicolumn{1}{c}{\textbf{\begin{tabular}[c]{@{}c@{}}Performance on\\ Dataset 1 (all)\end{tabular}}} \\ \hline
SVM & 62.1\% & 57.39\% & 60.43\% \\ \hline
k-NN & 69.2\% & 54.9\% & 66.28\% \\ \hline
DT & 68.03\% & 55.04\% & 64.16\% \\ \hline
RF & 70.47\% & 55.07\% & 66.96\% \\ \hline
MLP & 50.51\% & 49.62\% & 48.19\% \\ \hline
\end{tabular}%
% }
\end{table}

The performance of models trained on Dataset 1 varied from 53\% for the MLP model up to 58.88\% for the SVM, Table~\ref{tab:performance-dataset-1}, these results being in-line with state-of-the-art work by England et al.~\cite{england2019comparing} and Huang et al.~\cite{huang2019using}. Likewise, the models trained on the augmented Dataset 2 exhibit similar performance with the MLP architecture performing poorly with 49.62\% accuracy and SVM obtaining up to 57.39\% accuracy on the testing set, Table~\ref{tab:performance-dataset-2}.

\subsubsection{Evaluation on respective datasets:}
Indeed, the performance of the models on the respective testing sets is quite promising and substantially better than random choice ($\approx16.67\%$). However, we observe that the original version of the collected MetiTarski data (Dataset 1) is highly imbalanced (cf. Section~\ref{ssec:metitarski-datasets}). Prior research in domains such as computer vision and natural language processing has emphasized the negative impacts of an imbalanced dataset, demonstrating that the accuracy exhibited by models trained on imbalanced data could be \emph{deceiving} and hinder the capability of the model to learn meaningful information from the data. In most cases the model will exhibit a bias towards the label (class) which it has seen the most in the training set.

As such, it is interesting to investigate the performance of models trained on Dataset 1 (the original MetiTarski dataset) and the newly produced Dataset 2, with the latter being a superset of Dataset 1. As can be noticed in Table~\ref{tab:performance-dataset-1}, there is a significant drop in classification accuracy for all the models trained on Dataset 1, with more than 25\% drop in some cases. Models trained on the ``debiased'' Dataset 2 retain a good performance when evaluated on Dataset 1. We believe this is due to the training data being balanced and the model potentially seeing some of these samples during training, thus increasing its decision confidence. 

% !TEX root = ../main.tex
\section{Discussion and Future Work}
\label{sec:conclusions}

We have re-examined a classical dataset for ML-driven CAD variable ordering and observed issues of bias.
To address this, we have applied symmetry-based data augmentation to create a debiased version of the dataset\footnote{\url{https://github.com/coproof/variable-ordering-revisited}} and have shown this improves generalizability.
We believe that this phenomenon is quite general, and that debiasing
using formula symmetries should be a standard tool for applications of ML
in computer algebra, program verification, and other fields manipulating
mathematical formulas. We plan to extend this work into general-purpose
tools for ML-based variable ordering and apply it to many problem
domains (e.g., Gr\"obner bases, BDDs, SAT, etc.).

% !TEX root = ../main.tex
\subsubsection{Acknowledgements} This material is based upon work supported by the Defense Advanced Research Projects Agency (DARPA) under Contract No.\ HR00112290064. Any opinions, findings, and conclusions or recommendations expressed in this material are those of the author(s) and do not necessarily reflect the views of the United States Government or DARPA.
% </sections>
\bibliographystyle{splncs04}
\bibliography{main}

\begin{thebibliography}{10}
\providecommand{\url}[1]{\texttt{#1}}
\providecommand{\urlprefix}{URL }
\providecommand{\doi}[1]{https://doi.org/#1}

\bibitem{MetiTarski2010JAR}
Akbarpour, B., Paulson, L.C.: {MetiTarski}: An automatic theorem prover for
  real-valued special functions. Journal of Automated Reasoning  \textbf{44},
  175--205 (2010)

\bibitem{chen2020varord}
Chen, C., Zhu, Z., Chi, H.: Variable ordering selection for cylindrical
  algebraic decomposition with artificial neural networks. In: Bigatti, A.M.,
  Carette, J., Davenport, J.H., Joswig, M., de~Wolff, T. (eds.) Mathematical
  Software -- ICMS 2020. pp. 281--291. Springer International Publishing, Cham
  (2020)

\bibitem{collins1975cad}
Collins, G.E.: Quantifier elimination for real closed fields by cylindrical
  algebraic decompostion. In: Brakhage, H. (ed.) Automata Theory and Formal
  Languages. pp. 134--183. Springer Berlin Heidelberg, Berlin, Heidelberg
  (1975)

\bibitem{england2019comparing}
England, M., Florescu, D.: Comparing machine learning models to choose the
  variable ordering for cylindrical algebraic decomposition. In: International
  Conference on Intelligent Computer Mathematics. pp. 93--108. Springer (2019)

\bibitem{geirhos2018imagenettrained}
Geirhos, R., Rubisch, P., Michaelis, C., Bethge, M., Wichmann, F.A., Brendel,
  W.: Imagenet-trained {CNN}s are biased towards texture; increasing shape bias
  improves accuracy and robustness. In: International Conference on Learning
  Representations (2019), \url{https://openreview.net/forum?id=Bygh9j09KX}

\bibitem{geirhos2018generalisation}
Geirhos, R., Temme, C.R., Rauber, J., Sch{\"u}tt, H.H., Bethge, M., Wichmann,
  F.A.: Generalisation in humans and deep neural networks. Advances in neural
  information processing systems  \textbf{31} (2018)

\bibitem{huang2019using}
Huang, Z., England, M., Wilson, D.J., Bridge, J., Davenport, J.H., Paulson,
  L.C.: Using machine learning to improve cylindrical algebraic decomposition.
  Mathematics in Computer Science  \textbf{13}(4),  461--488 (2019)

\bibitem{kawaguchi2017generalization}
Kawaguchi, K., Kaelbling, L.P., Bengio, Y.: Generalization in deep learning.
  arXiv preprint arXiv:1710.05468  (2017)

\bibitem{Passmore2011PhDthesis}
Passmore, G.O.: Combined Decision Procedures for Nonlinear Arithmetics, Real
  and Complex. Ph.D. thesis, University of Edinburgh (2011)

\bibitem{passmore2012real}
Passmore, G.O., Paulson, L.C., Moura, L.d.: Real algebraic strategies for
  {MetiTarski} proofs. In: International Conference on Intelligent Computer
  Mathematics. pp. 358--370. Springer (2012)

\bibitem{Scarselli2009GNNs}
Scarselli, F., Gori, M., Tsoi, A.C., Hagenbuchner, M., Monfardini, G.: The
  graph neural network model. IEEE Transactions on Neural Networks
  \textbf{20}(1),  61--80 (2009). \doi{10.1109/TNN.2008.2005605}

\bibitem{vaswani2017attention}
Vaswani, A., Shazeer, N., Parmar, N., Uszkoreit, J., Jones, L., Gomez, A.N.,
  Kaiser, {\L}., Polosukhin, I.: Attention is all you need. Advances in neural
  information processing systems  \textbf{30} (2017)

\end{thebibliography}
\end{document}